**Title: Distortive Effects of Initial-Based Name Disambiguation on Measurements of Large-Scale Coauthorship Networks**

**Authors: Jinseok Kim and Jana Diesner**

**Author Information:**

Jinseok Kim
Graduate School of Library and Information Science, University of Illinois at Urbana-Champaign
501 E. Daniel St. Champaign, IL U.S.A. 61820
jkim362@illinois.edu

Jana Diesner
Graduate School of Library and Information Science, University of Illinois at Urbana-Champaign
501 E. Daniel St. Champaign, IL U.S.A. 61820
jdiesner@illinois.edu

**Corresponding Author:**
Jinseok Kim, jkim362@illinois.edu

## Abstract

Scholars have often relied on name initials to resolve name ambiguities in large-scale coauthorship network research. This approach bears the risk of incorrectly merging or splitting author identities. The use of initial-based disambiguation has been justified by the assumption that such errors would not affect research findings too much. This paper tests this assumption by analyzing coauthorship networks from five academic fields – biology, computer science, nanoscience, neuroscience, and physics – and an interdisciplinary journal, PNAS. Name instances in datasets of this study were disambiguated based on heuristics gained from previous algorithmic disambiguation solutions. We use disambiguated data as a proxy of ground-truth to test the performance of three types of initial-based disambiguation. Our results show that initial-based disambiguation can misrepresent statistical properties of coauthorship networks: it deflates the number of unique authors, number of component, average shortest paths, clustering coefficient, and assortativity, while it inflates average productivity, density, average coauthor number per author, and largest component size. Also, on average, more than half of top 10 productive or collaborative authors drop off the lists. Asian names were found to account for the majority of misidentification by initial-based disambiguation due to their common surname and given name initials.

## Introduction

Studies of large-scale[1] coauthorship networks have provided us with a bird-eye view of collaboration patterns in different fields. For example, Barabási and colleagues (2002) found that collaboration in mathematics and neuroscience is governed by preferential attachment: scholars choose to collaborate with others who already have many collaborators. Such a mechanism was shown to lead to a scale-free network, where the distribution of node degree follows a power law. Newman (2001, 2004) investigated coauthorship networks in biology, physics, computer science, and mathematics; showing that scholars in biology tend to have more papers and collaborators than those in the other fields considered. The studies also confirmed that coauthorship networks follow a power law distribution of node degree with cutoffs. A plethora of studies have followed to look into large-scale coauthorship networks to gain a detailed understanding of the power law regime in node degree distribution (e.g., Milojević, 2010), or to model the evolution of collaboration over time (e.g., Börner, Maru, & Goldstone, 2004), among other research questions.

Although not without exceptions (e.g., Franceschet, 2011; Strotmann & Zhao, 2012), many of large-scale coauthorship network studies have relied on given name initials to resolve name ambiguity during the pre-processing of bibliometric data. Some studies, for example, assume that if two name strings share a full surname and an initialized first name, they refer to the same author (e.g., Liben-Nowell & Kleinberg, 2007). This method regards 'Renear, A. H.' and 'Renear, A.' to refer to the same identity. Others look to the initials of first and middle names (e.g., Milojević, 2010). In this scheme, 'Renear, A. H.' and 'Renear, A.' represent two different authors because they do not share a middle name initial.

Scholars who use such an initial-based name disambiguation (IBD hereafter) have acknowledged that their method inevitably introduces errors of misidentification: two different authors can be regarded as one (i.e., merging) or one author can be divided into multiple identities (i.e., splitting). To justify the use of IBD, however, people have consistently argued that such errors are assumed to not much affect research findings (e.g., Barabási et al., 2002; Liben-Nowell & Kleinberg, 2007; Milojević, 2010; Newman, 2001, 2004). The problem here is that this argument has not been rigorously assessed. Considering the broad use of the IBD in bibliometrics (Strotman & Zhao, 2012) and the research findings of a high impact relying on this method[2], the investigation of the validity of this method is of great importance.

A few scholars have directly addressed this problem by comparing properties of two coauthorship networks: one which serves as a proxy for ground-truth network data and the other generated from the same data but disambiguated by name initials (Fegley & Torvik, 2013; Kim, Kim, & Diesner, 2014; Strotmann & Zhao, 2012). These studies found that the statistical properties of the networks are severely distorted by IBD. For example, when disambiguated by first and middle name initials, the number of unique authors in proxies of ground-truth datasets decreased by 31.23% (Fegley & Torvik, 2013) and 29.75% (Kim, Kim, & Diesner, 2014).

This study extends previous work by Fegley et al. (2013) and Kim et al. (2014) in that it compares properties of networks generated from a proxy for ground-truth data and by IBD versions of the same data. The contribution of this study is that it looks simultaneously into five different academic fields and an interdisciplinary journal, where IBD has been used for large-scale coauthorship network analysis. In addition, with author sizes ranging from 12,000 to 24,000, which are almost the same or lower numbers than those used in previous large-scale coauthorship network research, this study will provide an insight on the magnitude of the impact of IBD. Furthermore, this study investigates how much merging and splitting each affect the identification of a) actual individuals b) top authors in terms of productivity and popularity; with the latter being measured as number of coauthors. In the following section, we review how biases induced by IBD have been discussed in prior work.

## Background

Three kinds of IBD have been used in large-scale coauthorship network studies. In the explanations below, only name pairs with the same surname are supposed to be compared for IBD. So, when we say about names, we refer to given names, if otherwise defined. Also, a given name can have more than one token, where tokens are defined as space-separated sequences of one or more characters. For example, the given name of 'Renear, A. H.' has two name tokens, 'A.' and 'H.' The first approach of IBD would use the initial of the first name token (i.e., first initial based disambiguation, FD hereafter) (e.g., Bettencourt, Lobo, & Strumsky, 2007; Goyal, van der Leij, & Moraga-Gonzalez, 2006; Liben-Nowell & Kleinberg, 2007; Newman, 2001). If two name strings share an initial of first name tokens, then they are regarded to represent the same author. This method ignores the existence or differences in other name token initials. This can produce an error by merging two different authors into one (i.e., merging). For example, 'Renear, A. H.' can be merged with 'Renear, A. C.' into 'Renear, A.' – which all may be the same person or not.

The second method relies on initials of all name tokens (i.e., all initial based disambiguation, AD hereafter) (e.g. Barabási et al., 2002; Fiala, 2012; Milojević, 2010; Newman, 2001; Radicchi, Fortunato, Markines, & Vespignani, 2009; Rorissa & Yuan, 2012). According to this approach, only name instances sharing all the initials of name tokens represent the same author. For example, 'Renear, A. H.' and

'Renear, A. C.' would be different authors. This method is open to splitting errors. 'Renear, A. H.' is considered to be different from 'Renear, A.' even though they refer to the same author. Here, the second name token's initial in 'Renear, A.' may be missing due to inconsistent spelling. Some users of this method do not comment on their disambiguation strategy, or indicate that no disambiguation was performed and that names were used the way they occurred in a database or datasets (e.g., Börner, Maru, & Goldstone, 2004; Wagner & Leydesdorff, 2005). In other words, no actual disambiguation is performed. This approach is, however, the same as AD when datasets are supposed to entail a full surname and all available initial(s) of given name parts (e.g., WOS or SCOPUS datasets).

The third approach, a hybrid method, basically applies FD to name disambiguation. However, if a target name with a first name token's initial has two or more match candidates with different second name token's initials, then all of the target and candidate name strings are treated as referring to different identities (i.e., hybrid disambiguation, HD hereafter) (e.g., Milojević, 2013;Yoshikane, Nozawa, Shibui, & Suzuki, 2009). For example, if 'Renear, A.' has two match candidates, 'Renear, A. H.' and 'Renear, A. C.,' then those three names are considered to refer to different authors. If 'Renear, A.' has only one match candidate, 'Renear, A. H.' then those two names refer to the same author. This approach is open to both merging and splitting errors as it combines FD and AD.

Many scholars who use IBD have acknowledged that their approach can misidentify authors (e.g., Milojević, 2010; Newman, 2001), and showed different ways to correct these errors. In an attempt to increase the precision of name-identity matching, some scholars have leveraged affiliation information associated with author names (e.g., Acedo, Barroso, Casanueva, & Galan, 2006; Yoshikane et al., 2009). Due to the additional efforts required, this procedure is often restricted to a certain number of authors, e.g., the *N* most productive authors.

Alternatively, Newman (2001) simply assumed that the numbers of unique names identified by FD and AD correspond to the lower and upper bound of the "true" number of unique authors, respectively. Based on this assumption, he found that most statistical properties of coauthorship networks disambiguated by FD and AD showed errors of "an order of a few percent." This finding has been cited by many scholars to justify their use of IBD and to assume that their findings are not overly affected by their disambiguation methods (e.g., Barabási et al., 2002; Goyal et al., 2006; Liben-Nowell & Kleinberg, 2007; Milojević, 2010; Wagner & Leydesdorff, 2005; Yoshikane et al., 2009)[3].

The problem is that the assumption of supposedly negligible effect of IBD errors in large-scale coauthorship networks has not been tested in a rigorous way, including by those who have relied on it[4]. One way to test the performance of IBD would be to compare the properties of two coauthorship networks: one generated from the ground-truth data and the other disambiguated by initials. This idea is in line with the research tradition of social network scholars who have tested the robustness or stability of centrality measures when network data is flawed by removal or addition of nodes and links (e.g., Borgatti, Carley, & Krackhardt, 2006; Diesner & Carely, 2009; Frantz, Cataldo, & Carley, 2009).

A few scholars have applied this kind of testing method to large-scale coauthorship networks. For example, Fegley and Torvik (2013) compared two coauthorship networks: one generated from a proxy of ground-truth dataset disambiguated with advanced algorithms and the other by FD and AD. They found that IBD "dramatically" misrepresent network properties: e.g., 3.2 million unique authors identified by algorithms versus 1.56 million by FD and 2.18 million by AD. Another example is Kim et al. (2014)

where the same kind of distortion by IBD in the DBLP coauthorship network was detected using the same comparison strategy as Fegley and Torvik (2013) used.

Those studies, however, are limited in that the target datasets are exceptional in size: almost 3.2 million authors in MEDLINE and more than 775,000 in DBLP, respectively. It is still unknown whether such a distortive effect of IBD can be seen in coauthorship networks with authors of a few ten or hundred thousands, which constitute most of large-scale coauthorship networks in previous studies using initials for disambiguation. As the data size decreases, the errors induced by IBD may become reduced to a negligible extent. Another limitation is that different academic fields that follow different needs and traditions for coauthoring may show different levels of vulnerability to name ambiguity in coauthorship network data. As Strotmann and Zhao (2012) suggested, for example, errors from IBD may be more pronounced in a field such as nanoscience where scholars with Asian names are dominant. With regard to Kim and Diesner (2014), the accuracy of DBLP data has not been empirically tested, which may limit the robustness of the study.

To address these limitations, this study is designed to compare macroscopic properties of coauthorship networks generated from a proxy of ground-truth data and disambiguated by FD, AD, and HD, in five different academic fields and an interdisciplinary journal. The size of coauthorship networks are set to lie in the range between 12,000 and 24,000 in terms of the number of authors uniquely identified by FD, to approximate or be smaller than the size of selected large-scale coauthorship studies. Besides traditional measures such as node degree, clustering coefficient, or largest component size, this study introduces additional measures to gauge the effect of IBD: misidentification rate (Milojević, 2013) and top $k$-node comparison (Borgatti et al., 2006; Diesner & Carley, 2009; Frantz et al., 2009). In the following section, the selection of target field and acquisition of coauthorship network data is described in detail with the process of generating a proxy of ground-truth data, which is critical for impact evaluation of IBD.

## Method

*Data*

Data of large-scale coauthorship networks were obtained from the ISI Web of Science (WOS hereafter) for five selective academic fields (biology, computer science, nanoscience, neuroscience, and physics) and from one interdisciplinary journal (Proceedings of the National Academy of Sciences, PNAS hereafter). Sources, basic statistics, and references of selected large-scale coauthorship studies (> 10,000 estimated unique authors by IBD) conducted using IBD are shown in Table 1.

Table 1: Selected Large-Scale Coauthorship Network Research (Types of IBD are represented by FD = first initial based, AD = all initial based, and HD = hybrid method)

| Field | Research | Data Source | Year (Period) | No. of Journals | No. of Articles | No. of Estimated Unique Authors | Disambiguation Method |
|---|---|---|---|---|---|---|---|
| Biology (Biomedicine) | Newman (2001, 2004) | MEDLINE | 1995-1999 (5) | Not Specified | 2,163,923 | 1,520,251 | FD AD |
| Computer science | Fiala (2012) | WOS | 1996-2005 (10) | 426 | 205,780 | 187,016 | AD |

| | Yoshikane et al. (2009) | WOS | 1991-2005 (15) | 37 | 29,820 | 34,374 | HD |
|---|---|---|---|---|---|---|---|
| Nanoscience | Milojević (2010) | NanoBank | 2000-2004 (5) | 4,792 | 270,135 | 294,456 | AD |
| Neuroscience | Barabási et al. (2002) | Unknown | 1991-1998 (8) | "all relevant journals" | 210,750 | 209,293 | AD |
| Physics | Radicchi et al. (2009) | APS | 1893-2006 (114) | Physical Review Collection | 407,236 | 216,623 | AD |
| | Liben-Nowell et al. (2007) | arXiv | 1994-1999 (6) | Subfields in Physics | 35,555 | 23,589 | FD |
| Inter-disciplinary | Börner et al. (2004) | WOS | 1982-2001 (20) | 1 (PNAS) | 45,120 | 105,915 | AD |
| | Petersen et al. (2011) | WOS | 1958-2008 (51) | 6 (PNAS included) | 311,880 | 634,288 | AD |

Based on the proposition that the distortion of network measures due to name ambiguity increases with the size of networks (Fegley & Torvik, 2013), it is assumed that the effect of IBD observed in core journals within a limited time window (i.e., 1 year) can allow us to expect its effect to aggravate in a larger dataset encompassing a longer period. As each study varies in number of articles and estimated unique authors, the target number of articles is set to 5,000, which does not exceed the minimum size of articles in any of selected studies. Then, we selected the top 100 journals in four fields (biology, computer science, nanoscience, and neuroscience) according to the 5-year Impact Factor by the ISI Journal Citation Report 2012.

Next, we collected articles and pertinent metadata from the selected journals for the year 2012 from the WOS. We applied two restrictions to our search: 'article' (Document Types) and 'English' (Languages). We downloaded two versions of author names: one with 'full surname + initialized given names' and the other with 'full surname + full given names', if available. We started to download metadata of articles from journals with the high Impact Factor (in a descending order) in each target field. If the number of downloaded articles reached around 5,000, we stopped downloading. For physics, articles in *Physical Review B* were downloaded. For *PNAS*, all articles published in 2012 were retrieved. Single author papers and papers with 100 or more authors were excluded. In Table 2, the numbers of collected articles and journals are shown. The 'No. of Names' represent the number of name instances (not unique names) obtained for each dataset.

**Table 2: Summary of Downloaded Data**

| Fields | No. of Journals | No. of Articles | No. of Names |
|---|---|---|---|
| Biology | 50 | 5,569 | 27,679 |
| Computer Science | 40 | 5,235 | 19,178 |
| Nanoscience | 10 | 5,380 | 34,361 |
| Neuroscience | 22 | 5,314 | 35,172 |
| Physics | 1 | 5,417 | 27,084 |
| PNAS | 1 | 3,752 | 27,983 |
| Total | 124 | 30,667 | 171,457 |

*Name Disambiguation*

In order to test the effect of IBD on coauthorship networks, we need some ground-truth data where names are disambiguated correctly. Here, name disambiguation is a task to put relevant name instances into a cluster corresponding to a unique author identity based upon a similarity profile. For example, IBD compares name instances in terms of full surnames and initialized given names (i.e., similarity check via string matching). If they are found to share a full surname and first and/or middle name initials, they are grouped together (i.e., clustering).

The difficulty of name disambiguation has been intensively discussed among computer and information scientists, and various algorithmic solutions have been suggested. For comprehensive reviews on this topic, we refer readers to Smalheiser and Torvik (2009) and Ferreira, Gonçalves, and Laender (2012). We applied the findings from previous algorithmic disambiguation studies to our disambiguation process as follows.

In our disambiguation, four features per author – full name, names of coauthors, affiliation information, and email address - were used to produce a similarity profile of two name instance pairs. These features have been shown to be most effective in name disambiguation (Franceschet, 2011; Ley, 2009; Onodera et al., 2011; Torvik et al., 2005; Torvik & Smalheiser, 2009; Zhao & Strotmann, 2011). The full names of authors are used to select comparison pairs. We assign a unique numerical id to each name instance per dataset. Each name was first split into two parts: surname and given name. Then, any non-alphabetical characters such as a comma, a period, a hyphen, or an apostrophe, were deleted. Each surname and given name part was split based on spaces between tokens. Each token was assigned a position according to their order of appearance in the name string and only compared with the other name's token occupying the same position. For example, the name element 'Allen' in 'Lenear, Allen' is compared only to 'A.' in 'Lenear, A. P.'

In our disambiguation, name pairs are first put into string matching test of four steps to decide which pairs will be passed to the similarity calculation phase. Then, the comparable pairs are measured for similarity based on coauthor names, affiliation, and email address.

*Step1*: Name instances that match in every name token are compared to find cases of homonym (i.e., same names but different authors). If any two same name instances are identified to represent different authors, a unique id is attached to the end of the string of a name instance that appears behind the other in the order of unique-ids.

Steps two to four serve the identification of possible synonyms (i.e., different names but same authors).

*Step2*: Matching names that agree in the number of tokens, but where one or more tokens are not identical. If one of the unmatched tokens is a single character, then an initial match is performed: if the single character is the same as the first letter of the token from the other name, then the token pair is regarded as being an 'initialized match'. This does not apply to surname token comparison. If all of the unmatched tokens are initialized matches, we pass the pair to a similarity calculation. For example, 'Smith, John Loy' would be considered for similarity check with 'Smith, John L.' and 'Smith, J. L.'

*Step3*: Trying to find match candidates that differ in the number of tokens. If all tokens in a shorter name find a match in full string format or an 'initialized match' with tokens in a longer name, the two names are considered for similarity calculation. For example, 'Smith, John' and 'Smith, J. L.' are compared for similarity.

*Step4*: This step is illustrated in the Table 3, which looks into the name element matches with an edit distance of one, partial, or nick name matches[5], and the case of permutated tokens (Ley, 2009; Torvik & Smalheiser, 2009). Here, the matching scheme of Case 1 and Case 3 shown in the table also apply to surname comparison. For Case 4, all name tokens are compared for full or initial matches regardless of their position in each name string and of whether they are in surnames or given names. For Case 3, names with an Asian surname were not considered for matching because many of them differ in just one character (e.g., Liu, John. vs. Li, John.)[6].

Table 3: Selection of match candidates for synonyms

| Case | Examples from Torvik & Smalheiser (2009) |
|---|---|
| 1 | different only with/without a space<br>e.g., jean francois vs. jeanfrancois |
| 2 | different with a partial name or a nickname<br>e.g., zak vs zakaria<br>e.g., dave vs. david |
| 3 | different with one alphabetical character<br>e.g., bjoern vs. bjorn<br>e.g., bjoern vs. bjaern |
| 4 | different with permutated name tokens<br>e.g., kim, jin vs. jin, kim<br>e.g., garcia-moreno, fernando vs. moreno, fernando garcia |

In each of steps described above, selected name pairs are compared for similarity in terms of coauthors, affiliation such as a university address, and an email address, if any. The similarity calculation procedure is detailed as follows:

*Coauthor Similarity*: if two name instances share one or more coauthor names in a full given name format, then the pair is given a coauthor similarity value of 1.0 per match. If either or both of their coauthor names come with initialized given names and match in their initialized format (e.g., 'Renear, Allen' vs 'Renear, A.'), then the coauthor similarity value of the target name instances is 0.3 per match.

*Affiliation Similarity*: before calculating the affiliation similarity, a stoplist is applied to each affiliation name for the 20 most frequent words in each dataset. As the vocabulary in affiliation information in our dataset is controlled by the WOS, stemming was not performed. The ratio of matching words between two affiliation names over the number of words in a shorter affiliation name is set as the affiliation similarity value. For example, let's assume that the affiliation of name A has five words and the affiliation of name B has seven words. If they share four words, then the affiliation similarity is 4/5 = 0.8. If affiliation names share a zip-code (here, a string of numbers with 4 or more numeric characters), then the zip-code match adds 0.5 to the similarity score. If a name instance has two or more affiliation name, each one is compared and the maximum score is chosen[7].

*Email similarity*: if two name instances have email addresses that match in a full string except domain address part (e.g., @gmail.com), then the pair is given a binary similarity value of 1 for match and 0 for non-match.

If the sum of the similarity scores for coauthor, affiliation and email is over 0.50 (homonyms) or 0.75 (synonyms), then the target name pairs are assumed to refer to the same author. As the purpose of our disambiguation is to obtain disambiguated datasets with a high accuracy rather than to propose a good-performing algorithm, name pairs with similarity scores between 0.40 and 0.50 (homonyms) and between 0.40 and 0.75 (synonyms) are passed to manual checking[8]. All these parameters are empirically set by researchers. During the comparison of name pairs, transitivity violations may occur (Torvik & Smalheiser, 2009). For example, let's assume that name A is matched with name B and also with name C to represent the same author. If name B and C are found to mismatch, then the transitivity is violated. In every step, such violations are corrected if applicable.

*Evaluation of Disambiguation*

Authorship datasets disambiguated by our approach inevitably contain identification errors. The performance of our disambiguation method can be evaluated against a sample of ground-truth data that are manually disambiguated and can be measured by K metric and cluster F1 (Cota et al., 2010; Pereira et al., 2009). We believe these two metrics are appropriate to evaluate the performance of our disambiguation because in our study each author identity is represented by a cluster of name instances that are judged to refer to the author based on similarity of coauthor, affiliation, and email address.

*K-metric*: The K-metric is the geometric mean between two metrics: average cluster purity (ACP) and average author purity (AAP).

$$K = \sqrt{ACP \times AAP}$$

The equations to calculate ACP and AAP are shown below.

$$ACP = \frac{1}{N} \sum_{i=1}^{q} \sum_{j=1}^{R} \frac{n_{ij}^2}{n_i}$$

$$AAP = \frac{1}{N} \sum_{j=1}^{R} \sum_{i=1}^{q} \frac{n_{ij}^2}{n_j}$$

Here, *N* is the total number of name instances that are subject to disambiguation; *R* is the number of clusters generated by manual disambiguation (i.e., ground-truth clusters); *q* is the number of clusters generated by algorithmic disambiguation (i.e., our disambiguation); $n_{ij}$ is the total number of elements of cluster $i$ in $q$ belonging to the cluster $j$ in $R$; $n_i$ and $n_j$ represent total numbers of elements in the cluster $i$ and $j$, respectively.

ACP measures the purity of the clusters generated by algorithmic disambiguation: if all clusters contain only the correct name instances belonging to the same identities, then the ACP value will be 1. In other words, the value decreases if clusters include merged identities. Meanwhile, AAP measures the fragmentation of algorithmically generated clusters: if each cluster has a low proportion of name instances that should belong to this cluster but are not included in it, the value becomes closer to 1. In other words, this indicates how many identities are split into separate clusters.

*Cluster F1*: This is a harmonic mean of cluster precision and cluster recall. Cluster precision (*c*P) is calculated as *c*P = A/B, while cluster recall (*c*R) is calculated as *c*R = A/C. Here, A is the number of correct clusters, while B is the number of clusters generated by algorithmic disambiguation. C is the number of clusters generate from manual disambiguation. From these two metrics, the *c*F1 is defined as follows:

$$cF1 = \frac{2 \times cP \times cR}{(cP + cR)}$$

Many studies in algorithmic name disambiguation evaluate their methods against most ambiguous name groups such as Wang, S. or Kim, J. (e.g., Cota et al., 2010; Ferreira et al., 2012). Following this convention, we selected name instances from the 10 most ambiguous name groups in terms of shared first name initial in each dataset, and manually disambiguated them by leveraging information from email addresses, affiliation, coauthor names, personal CVs, and institution homepages associated with each name. Despite all the efforts, some names were impossible to disambiguate. In such cases, we assumed that each name represents different authors.

In Table 4, the evaluation results are reported along with the number of name instances and unique authors identified by manual disambiguation for each field. AAP (0.981 ~ 1.000) is higher than ACP across the fields, indicating that splitting of identities happens less often than merging. Our disambiguation led to various moderate degrees of identity merging (ACP: 0.965 ~ 0.997). Cluster F1 also shows that most of the name instances are clustered correctly. Based on the overall results from K-metric and cluster F1, we believe that the disambiguated datasets can serve as a *proxy* for ground-truth datasets, even though they are not perfectly disambiguated.

**Table 4: Evaluation of Our Disambiguation by K-Metric and Cluster F1**

| Fields | No. of Name Instances (No. of Unique Authors) | K-metric | | | Cluster F1 | | |
|---|---|---|---|---|---|---|---|
| | | ACP | AAP | K | *c*P | *c*R | *c*F1 |
| Biology | 288 (261) | 0.976 | 1.000 | 0.988 | 0.98 | 0.95 | 0.97 |
| Computer Science | 384 (285) | 0.997 | 1.000 | 0.999 | 1.00 | 0.99 | 0.99 |
| Nanoscience | 1,040 (724) | 0.965 | 0.998 | 0.981 | 0.98 | 0.95 | 0.96 |
| Neuroscience | 262 (230) | 0.972 | 1.000 | 0.986 | 0.98 | 0.95 | 0.97 |
| Physics | 317 (166) | 0.974 | 0.981 | 0.977 | 0.93 | 0.92 | 0.92 |
| PNAS | 319 (300) | 0.994 | 1.000 | 0.997 | 0.99 | 0.99 | 0.99 |

*Generating Comparison Datasets*

We applied three types of IBD, i.e., FD, AD, and HD, as described in Milojević (2013) and Newman (2001), to the abovementioned datasets. We use the author names in the format of full surname and initialized given name(s) as provided by the WOS. Thus, each authorship dataset has four corresponding datasets: a dataset disambiguated (1) by our semi-automatic algorithm, (2) FD, (3) AD, and (4) HD.

*Measurements*

This study employs 11 measurements to test the performance of IBD. They are selected for a comparison purpose as many previous large-scale coauthorship studies have used them. If a metric has more than one way to be calculated, we follow the approach taken in previous coauthorship studies. Network measurements were calculated in *Pajek* (de Nooy, Mrvar, & Batagelj, 2011) and *i*graph (Csardi & Nepusz, 2006).

*Unique Author*: This represents the number of author identities (i.e., clusters) uniquely identified by our disambiguation as well as the three types of IBD.

*M-Rate* (Milojević, 2013): This calculates how many unique authors in the proxy of ground-truth data have been misidentified by IBD. A unique author is misidentified if his/her identity is merged with other identity and/or is split into two or more identities. Thus, the misidentification rate of each IBD is expressed as the ratio of unique authors (= clusters) in the proxy of ground-truth data whose identity is merged and/or split by IBD[9].

*Productivity*: The number of papers per author. The merged or split author identities directly affect productivity as they can inflate or deflate the number of publications of each author.

*Degree Centrality*: Any two scholars are connected via a coauthorship relation, i.e. an edge, if they have coauthored a paper. Here, only the existence of collaboration among authors matters: the frequency of collaboration is ignored (Barabási et al., 2002; Newman, 2001). Therefore, the degree of an author represents the number of her unique collaborators.

*Number of Edges*: An edge represents the existence of collaboration between two authors. The number of edges is the total number of unique edges in a coauthorship network.

*Density*: Density measures the proportion of the number of actual edges over the number of possible edges.

*Component*: A component in a coauthor network is a maximal set of authors where each author can reach every other by several steps. Scholars have been interested in the size (and its ratio) of the largest component because it can tell us about the degree to which a field of science is well connected (i.e., mature) or vulnerable to fragmentation (Newman, 2001).

*Shortest Path*: The shortest path, or the geodesic, between any two authors is the minimum number of steps between them within a component. In this study, the average shortest paths of all reachable authors in each dataset (Brandes, 2001) are reported. Only the lengths of the existing paths were averaged.

*Clustering Coefficient*: The clustering coefficient measures the average fraction of pairs of an author's collaborators who have also collaborated with one another (Newman, 2001). This represents the degree of triadic closure among authors.

A*ssortativity*: This measures the extent to which authors collaborate with others who are similar to them in terms of degree centrality. In this study, assortativity is calculated as "the Pearson correlation coefficient of the degrees at either ends of an edge" between any two authors (Newman, 2002).

*Top k Authors:* Social network scholars have tested the robustness of network measurements by looking into the difference of top k nodes found in the ground-truth network data and those found in the flawed network data (Borgatti et al., 2006; Diesner & Carley, 2009; Frantz et al., 2009). We apply this approach to identify the sets of the top 1, 10, and 20 authors in terms of productivity and degree centrality.

**Results**

*Number of unique authors*

In Table 5, the number of unique authors per dataset is shown along with its change rate in comparison to the ground-truth proxy (PGT hereafter). Overall, all of three IBD methods underestimated the size of unique authors. For example, in nanoscience, FD reduced the number of unique authors by 31%. Overall, FD tends to underestimate more than AD and HD did. The difference between FD and AD estimations is often larger than a few percent, contrary to Newman (2001)'s proposition.

Table 5: Number of Unique Authors Identified by PGT (Proxy of ground-truth), FD, AD, and HD

| Field | PGT | FD | Change (%) | AD | Change (%) | HD | Change (%) |
|---|---|---|---|---|---|---|---|
| Biology | 24,639 | 21,520 | -12.66 | 23,681 | -3.89 | 23,136 | -6.10 |
| Computer Science | 15,932 | 12,646 | -20.63 | 14,744 | -7.46 | 14,367 | -9.82 |
| Nanoscience | 25,535 | 17,553 | -31.26 | 21,904 | -14.22 | 21,287 | -16.64 |
| Neuroscience | 27,281 | 24,163 | -11.43 | 26,476 | -2.95 | 25,683 | -5.86 |
| Physics | 17,496 | 15,225 | -12.98 | 16,543 | -5.45 | 16,156 | -7.66 |
| PNAS | 25,915 | 22,157 | -14.50 | 24,641 | -4.92 | 24,152 | -6.80 |

The most noticeable finding here is, however, the off-upper-bound phenomenon. In the Table 5, all the numbers of unique authors in the proxy of ground-truth data are found off the upper bounds, i.e., where the upper bound is largest numbers of unique authors estimated by IBD. This finding contrasts with Newman (2001) and other's assumption that AD and FD can provide the upper and lower bounds of the number of unique authors. Previous studies (Fegley & Torvik, 2013; Kim, Kim, & Diesner, 2014) also found the off-upper-bound phenomenon in MEDLINE and DBLP datasets, respectively.

*Misidentification Rate (M-Rate)*

The underestimation of the number of unique authors indicates that IBD mostly merges author identities where there could be certain levels of splitting. Table 6 shows the M-Rates, ratios of misidentified (i.e., merged and/or split) authors over the proxy of ground-truth data, and detailed ratios of merging, splitting, and merging & splitting cases in each dataset. The ratio of author identities compromised by IBD does not seem to be trivial. For example, in nanoscience, about 39.17% of all unique authors in the proxy of ground-truth data have been merged and/or split by FD. At a minimum, 8.37%% of the unique authors are misidentified by AD in neuroscience. AD performed consistently better than other two methods, which is contrary to the finding of Milojević (2013) that HD often approximates the *hypothesized* ground-truth data better than FD and AD, and FD is superior to AD in detecting true author identities. As expected, merging is more prevalent in FD than in AD and HD, and splitting errors happened more often in AD than in FD and HD. However, merging is dominant in all IBD methods, which seem to contribute to the decrease of uniquely identified authors by each of the IBD compared to the proxy of ground-truth.

Table 6: Ratio of Misidentified Authors in PGT (proxy of ground-truth) data by IBD

| Field | No. of Unique Authors in PGT | FD M-Rate | AD M-Rate | HD M-Rate |
|---|---|---|---|---|
| Biology | 24,639 | 18.90% (s: 0.34%, m: 99.63%, s&m: 0.02%) | 7.81% (s: 5.93%, m: 93.71%, s&m: 0.36%) | 10.75% (s: 0.87%, m: 98.94%, s&m: 0.19%) |
| Computer Science | 15,932 | 27.60% (s: 0.09%, m: 99.91%, s&m: 0.00%) | 13.19% (s: 3.95%, m: 95.76%, s&m: 0.29%) | 16.09% (s: 0.59%, m: 99.22%, s&m: 0.20%) |
| Nanoscience | 25,535 | 39.17% (s: 0.20%, m: 99.74%, s&m: 0.06%) | 22.77% (s: 2.80%, m: 96.11%, s&m: 1.08%) | 25.43% (s: 0.55%, m: 98.61%, s&m: 0.83%) |
| Neuroscience | 27,281 | 17.68% (s: 0.37%, m: 99.61%, s&m: 0.02%) | 8.37% (s: 14.06%, m: 84.98%, s&m: 0.96%) | 10.62% (s: 1.28%, m: 98.34%, s&m: 0.38%) |
| Physics | 17,496 | 19.60% (s: 0.20%, m: 99.77%, s&m: 0.03%) | 11.98% (s: 6.68%, m: 91.36%, s&m: 1.96%) | 13.67% (s: 0.59%, m: 98.33%, s&m: 1.09%) |
| PNAS | 25,915 | 21.02% (s: 0.06%, m: 99.94%, s&m: 0.00%) | 8.49% (s: 1.77%, m: 97.91%, s&m: 0.32%) | 11.50% (s: 0.13%, m: 99.70%, s&m: 0.17%) |

*Productivity and Degree Distributions*

To illustrate the effect of merged and/or split author identities on statistical properties, the productivity and number of coauthors (i.e., degree centrality) distribution are shown in Figures 1 and 2. For each cumulative log-log plot, we show the distribution for each of the three disambiguation strategies: PGT (black circles), AD (blue crosses) and FD (red dots). The distribution from HD is omitted for visual simplicity because HD is found to overlap with AD on the most part of its curve in all datasets.

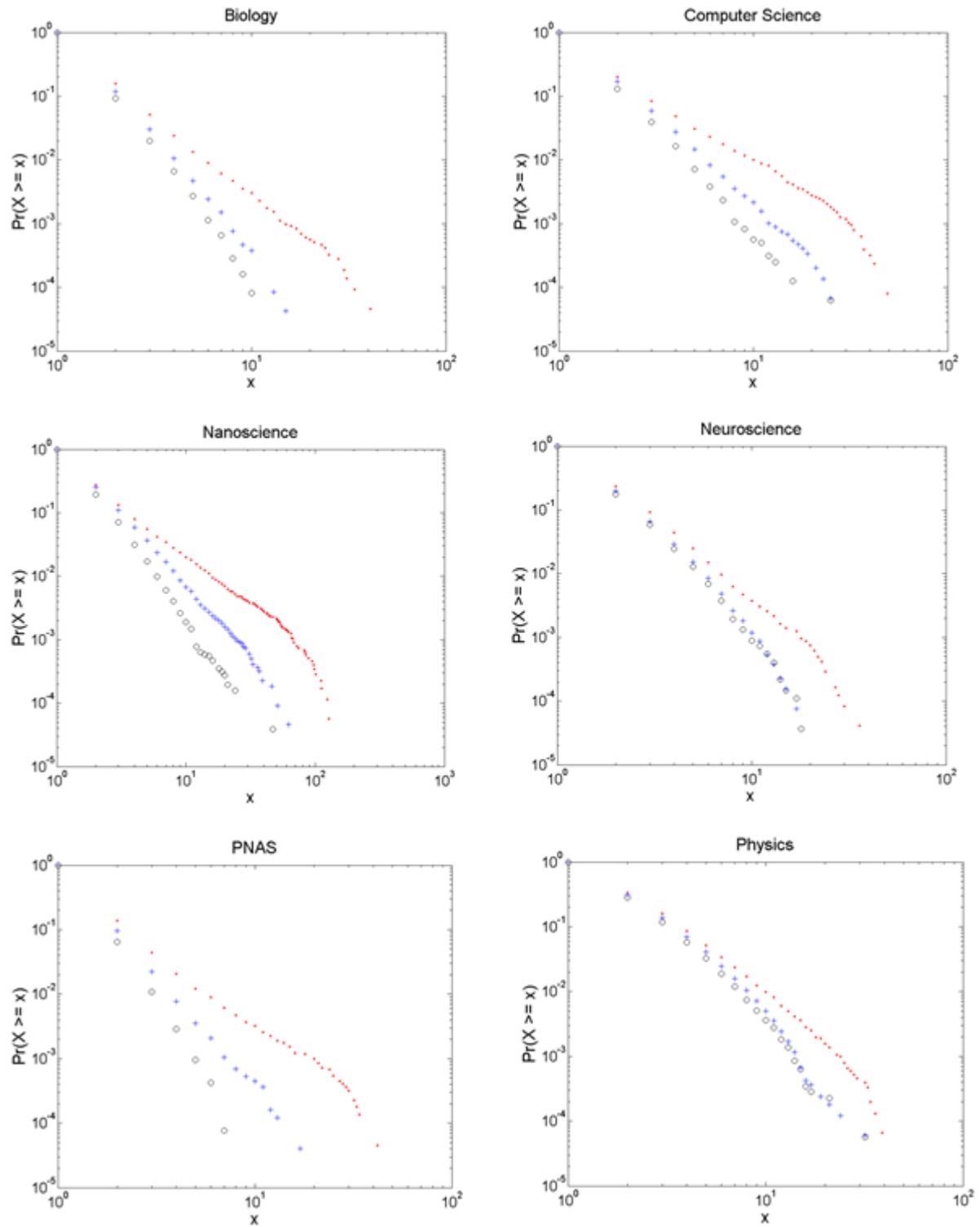

Figure 1: Cumulative Log-Log Plot of Productivity Distribution (black circles for PGT, blue crosses for AD, and red dots for FD)

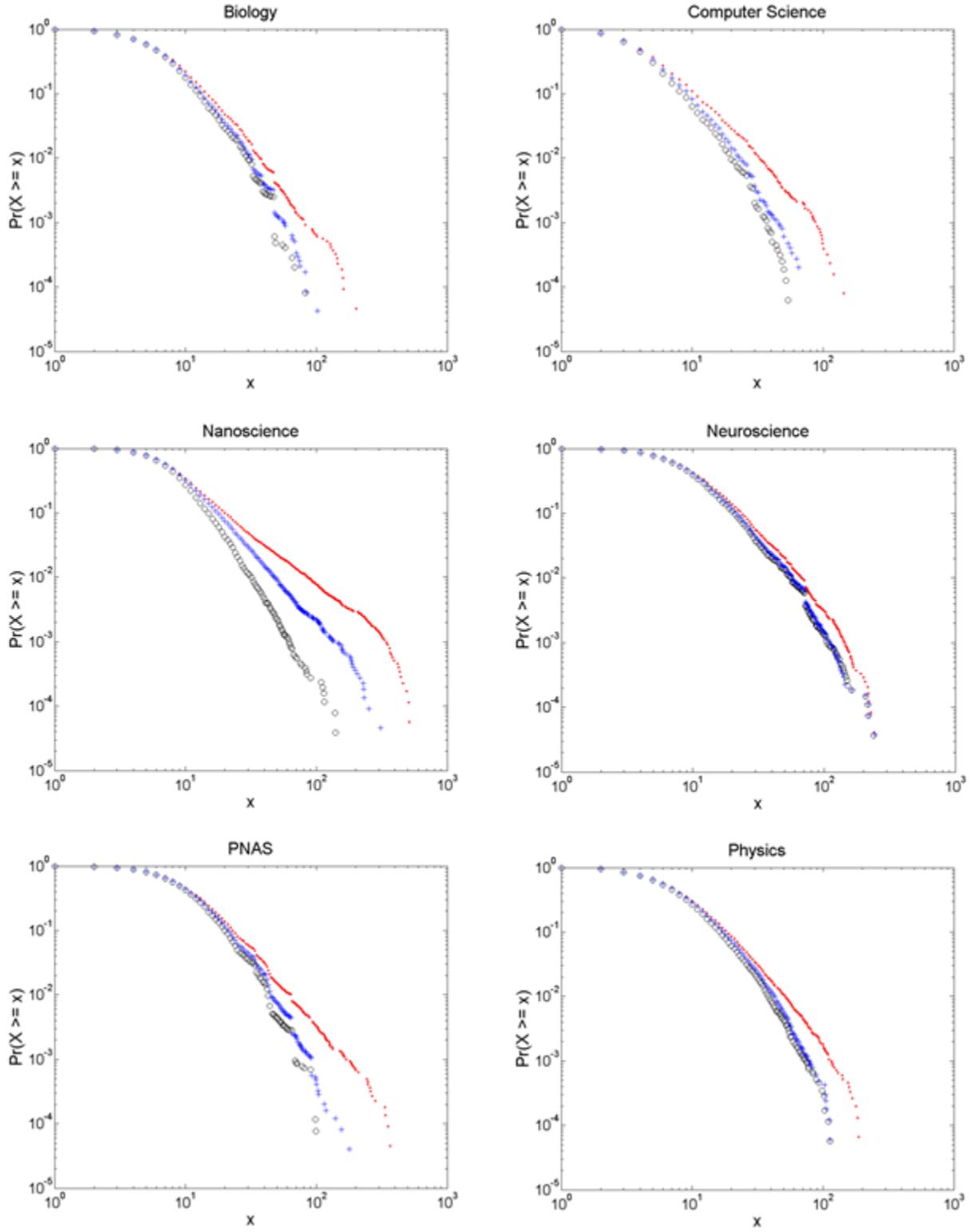

Figure 2: Cumulative Log-Log Plot of Degree Distribution (black circles for PGT, blue crosses for AD, and red dots for FD)

The distribution curves of the proxy of ground-truth datasets show more curvature downward compared to those for FD (in all cases) and AD (except in neuroscience and physics). This means that for a given value (x) of productivity or number of coauthors, the proportion of authors who have the given value or above (X >= x) increases by IBD, and this happens for the majority of x values. In other words, the blue cross and red dot curves positioned above the black circle curves mean that productivity and coauthor size distributions are distorted. IBD methods create merged identities including inflated values of productivity and number of coauthors, which pushes the curves both upward and right. Moreover, in many cases, the curves from AD and FD in Figure 2 seem to be straighter with a lower slope than those from the proxy of ground-truth data, which may fit better than the PGT data curves into the so-called power-law distribution. In an analysis of coauthor network of scholars in MEDLINE for the 2003-2007 periods, Fegley and Torvik (2013, Figure 8) evidenced the same finding: the distribution of the number of coauthors from the algorithmic disambiguation shows much more curvature than one from an all-initial-based disambiguation. Kim, Kim, and Diesner (2014) also reported the same finding in DBLP data.

In neuroscience and physics, the gaps between AD and the proxy of ground-truth curves in two figures seem to be smaller compared to other fields. However, one limitation here is that the AD curves come from the data that include a substantial number of misidentified authors (8.37 % in neuroscience and 11.98% in physics). In other words, even though the AD curves seem to be close to the PGT curves, each distribution is generated from different sets of authors. Thus, a further investigation seems to be needed to understand if and why the AD curves approximate true distribution of the productivity and the collaborator numbers in two fields.

*Statistical Properties*

In the Table 7, eight statistical properties of each data are reported. Except for average productivity, they address network structure: number of edges, density, average degree, transitivity, average shortest paths, assortativity, and the number of components and size of the largest component.

The various disambiguation strategies have a small impact on the number of edges: on average, FD and HD reduced the number by 1.20% and 0.17%, respectively, while AD increased it by 0.11%. This suggests that merged author nodes typically have distinct sets of coauthors. If two merged author identities have coauthors that are also merged due to name ambiguity, then the ties between each author and coauthor would also be merged, leading to the decrease of ties.

However, this minor change comes along with more drastic changes for the other metrics: density showed a tendency to increase by IBD. The network densities of PGT network increased on average 47.52% by FD, 14.93% by AD, and 20.72% by HD. In general, density tends to decrease with increasing network size. However, the observed increases are much stronger than the change in the number of edges. Since the number of edges remains fairly stable as reported above, we assume the decrease in the number of unique authors by IBD to mostly affect the increase of densities.

The average productivity of authors tended to increase due to IBD (on average 21.96% by FD, 7.26% by AD, and 9.87% by HD). Its standard deviations also increased. This means IBD creates more outliers than would be expected in the proxy of ground-truth datasets. The average degree also increased in all datasets

(on average 20.23% by FD, 7.12% by AD, and 9.63% by HD); also with increased standard deviations. The increased values of productivity and degree by IBD were not unexpected: it is logical that authors who are merged by IBD get associated with more publications (some of which were truly authored by others) while the total number of unique authors decreases. Likewise, merged authors get connected to others who are not their actual collaborators, contributing to the inflation in the average number of coauthors.

The size of largest components was increased by on average 37.26 %$p$ by FD, 24.08%$p$ by AD, and 30.11%$p$ by HD, while the number of components decreased. As authors are merged into other identities, they attach their local networks to others, thus increasing the size of their component. Some fields showed noticeable increases, which might lead to different interpretation of the coauthorship network structure. For example, in biology, the largest component in FD data contained 49.25% of all authors in the FD data, while the largest component of PGT data contained only 0.72% of authors in the PGT data. We observed the same finding for PNAS. From the perspective of FD, the collaboration network in biology and PNAS dataset is way more inter-connected and mature than it actually is.

The overall decrease of the average shortest paths (on average 39.93% by FD, 23.04% by AD, and 25.42% by HD; biology was excluded from this percentage calculation) might be explained similarly to the increase of the largest component size. As more authors got integrated into larger components, merged authors seem to work as a bridge connecting the previously unreachable authors and to provide shortcuts for connecting them. For example, in computer science, authors in PGT data were found to reach one another within on average 11.66 steps of collaboration partners, while those in FD data are reachable with only on average 5.32 steps. One exception is biology, where the average shortest path is 2.36, and increased by IBD. The coauthorship network for biology is not well connected (the largest component has 178 nodes among 24,639 nodes), which contributes to the low average shortest paths.

IBD leads to decreases in the clustering coefficients (a.k.a. transitivity) (on average 39.33% by FD, 14.52% by AD, and 16.89% by HD). This can also be explained mainly by the merging effect. We calculated the clustering coefficient as the fraction of actual triangles over the triples connected with two edges (possible triangles). The merging of author identities in coauthorship network increases the number of connected triples (denominator), while the number of triangles (numerator) does not increase at a corresponding rate. In other words, when two authors come to be connected through a merged author (i.e., forms a connected triple), those two authors are not expected to have collaborated actually (i.e., fails to form a triangle). If we rely on the intuitive interpretation of clustering coefficient (Franceschet, 2011), authors in network datasets disambiguated by IBD are depicted as being less inclined to collaborate with others sharing a common collaborator than they actually are.

Assortativity decreased by on average 53.22% by FD, 25.89% by AD, and 25.96% by HD. This indicates that IBD make authors appear to collaborate less with others who are similar to them in terms of the number of coauthors. For example, in PNAS, scholars in the PGT data showed a strong tendency (0.829) to collaborate with other collaborative scholars, while solitary authors tend to work with other solitary authors. Such a collaboration structure is shown to nearly disappear (0.160) in the FD data or to be weakened (0.528 or 0.543) in the AD or HD data.

In sum, IBD has a slight quantitative impact on the number of edges, and leads to strong increases (density, degree centrality, and largest component size) and decreases (number of components, average shortest paths, transitivity, and assortativity) in several key network metrics. Overall, changes were more pronounced by FD, followed by HD, and by AD. The key take away and implication here is that the false impressions we can get of a network due to incorrect disambiguation can lead to false theories and a biased understanding of the structure, functioning and dynamics of coauthoring; and cause unjustified actions, such as pertinent policy and funding decisions.

Table 7: Overview of Statistical Properties of Coauthorship Networks per Disambiguation Method

| Field & Method | | No. of Edges | Density | Avg. Productivity (SD) | Avg. Degree (SD) | No. of Components (Ratio of Largest Component Size) | Avg. Shortest Path | Transitivity | Assortativity |
|---|---|---|---|---|---|---|---|---|---|
| Biology | PGT | 80,253 | 2.64E-04 | 1.12 (0.46) | 6.51 (5.48) | 3,881 (0.72%) | 2.36 | 0.908 | 0.836 |
| | FD | 79,686 | 3.44E-04 | 1.29 (1.12) | 7.41 (7.75) | 2,088 (49.25%) | 7.00 | 0.638 | 0.288 |
| | AD | 80,336 | 2.87E-04 | 1.17 (0.58) | 6.78 (5.97) | 3,028 (23.68%) | 8.94 | 0.834 | 0.649 |
| | HD | 80,199 | 3.00E-04 | 1.20 (0.62) | 6.93 (6.11) | 2,646 (34.37%) | 9.49 | 0.812 | 0.636 |
| Computer Science | PGT | 33,391 | 2.63E-04 | 1.20 (0.70) | 4.19 (3.76) | 3,160 (9.95%) | 11.66 | 0.794 | 0.637 |
| | FD | 33,138 | 4.14E-04 | 1.52 (2.05) | 5.24 (6.74) | 1,690 (47.21%) | 5.32 | 0.415 | 0.231 |
| | AD | 33,456 | 3.07E-04 | 1.30 (0.98) | 4.53 (4.37) | 2,288 (36.93%) | 7.34 | 0.673 | 0.400 |
| | HD | 33,364 | 3.23E-04 | 1.33 (1.02) | 4.64 (4.50) | 2,069 (41.32%) | 7.23 | 0.651 | 0.399 |
| Nanoscience | PGT | 105,286 | 3.23E-04 | 1.35 (1.04) | 8.25 (6.14) | 1,677 (54.37%) | 8.40 | 0.647 | 0.174 |
| | FD | 101,414 | 6.58E-04 | 1.96 (4.34) | 11.56 (21.39) | 552 (82.14%) | 4.47 | 0.191 | 0.102 |
| | AD | 105,250 | 4.38E-04 | 1.57 (1.80) | 9.60 (10.52) | 806 (78.04%) | 5.26 | 0.383 | 0.092 |
| | HD | 104,969 | 4.63E-04 | 1.61 (1.85) | 9.86 (10.76) | 695 (80.87%) | 5.22 | 0.375 | 0.097 |
| Neuroscience | PGT | 143,478 | 3.86E-04 | 1.29 (0.83) | 10.52 (10.51) | 1,584 (60.68%) | 8.40 | 0.718 | 0.530 |
| | FD | 142,746 | 4.89E-04 | 1.46 (1.30) | 11.82 (12.94) | 676 (86.10%) | 5.54 | 0.577 | 0.327 |
| | AD | 143,986 | 4.10E-04 | 1.33 (0.90) | 10.86 (10.90) | 1,138 (75.42%) | 7.31 | 0.690 | 0.499 |
| | HD | 143,296 | 4.35E-04 | 1.37 (0.95) | 11.16 (11.26) | 939 (80.45%) | 6.75 | 0.668 | 0.491 |
| Physics | PGT | 69,768 | 4.56E-04 | 1.55 (1.28) | 7.98 (7.58) | 1,164 (66.76%) | 7.28 | 0.553 | 0.290 |
| | FD | 69,169 | 5.97E-04 | 1.78 (1.89) | 9.09 (10.38) | 656 (82.10%) | 5.43 | 0.397 | 0.204 |
| | AD | 69,809 | 5.10E-04 | 1.64 (1.41) | 8.44 (8.23) | 834 (78.14%) | 6.36 | 0.507 | 0.266 |
| | HD | 69,571 | 5.33E-04 | 1.68 (1.45) | 8.61 (8.43) | 755 (80.30%) | 6.20 | 0.495 | 0.271 |
| PNAS | PGT | 134,972 | 4.02E-04 | 1.08 (0.34) | 10.42 (8.28) | 2,554 (4.89%) | 8.04 | 0.914 | 0.829 |
| | FD | 134,064 | 5.47E-04 | 1.26 (1.20) | 12.10 (13.76) | 969 (74.17%) | 4.90 | 0.547 | 0.160 |
| | AD | 134,962 | 4.44E-04 | 1.14 (0.53) | 10.95 (9.35) | 1,679 (49.64%) | 6.82 | 0.816 | 0.543 |
| | HD | 134,777 | 4.62E-04 | 1.16 (0.57) | 11.16 (9.59) | 1,408 (60.76%) | 6.69 | 0.793 | 0.528 |

*Top k Authors*

Another argument in support of IBD is that it can successfully detect the most prolific or well-connected scholars (Milojević, 2013). These people are often considered as prominent and influential in knowledge production and dissemination. In order to see if disambiguation affects author rankings in terms of productivity and number of coauthors, we compared the top 1, 10, and 20 authors. During this task, we found that both proxy of ground-truth and IBD datasets produced many authors with tied rankings between the top 10 and the top 20, and the number of tied authors differed greatly depending on fields. Thus, for top 20 authors, instead of specifying the ranking level of interest, we find all the authors in each dataset who have publications or collaborators beyond certain thresholds that enable us to list at least top 20 authors.

Tables 8 and 9 summarize the number of top ranking authors in terms of productivity and number of coauthors. For example, in Table 8, biology has 28 authors in the proxy of ground-truth who have 6 or more publications. If the threshold was set to 7, the number of authors dropped to sixteen, while when set to 5, it increased to 67. According to IBD, for example, the number of authors who have the same threshold productivity in biology increased up to 192 (585%). This kind of inflation happened across all fields. Tables 8 and 9 also tell us how many of the top 1 and 10 authors in each dataset have stayed or dropped off lists by IBD. These changes seem not negligible.

**Table 8: Top 1, 10, and 20 Authors by Productivity**

| Field | No. of Top 20 in PGT (Threshold) | No. of Authors on or above Threshold in PGT | | | No. of Scholars who Stay in Top 10 | | | Change of Top 1 (Y/N) | | |
|---|---|---|---|---|---|---|---|---|---|---|
| | | FD | AD | HD | FD | AD | HD | FD | AD | HD |
| Biology | 28 (6) | 192 | 57 | 65 | 0 | 3 | 3 | Y | Y | Y |
| CS | 37 (7) | 225 | 81 | 91 | 2 | 5 | 5 | Y | N | N |
| Nanoscience | 20 (12) | 274 | 96 | 100 | 2 | 5 | 5 | Y | Y | Y |
| Neuroscience | 20 (11) | 74 | 23 | 25 | 0 | 6 | 8 | Y | Y | Y |
| Physics | 24 (13) | 76 | 28 | 28 | 1 | 6 | 7 | Y | N | N |
| PNAS | 25 (5) | 267 | 88 | 98 | 0 | 0 | 0 | Y | Y | Y |

**Table 9: Top 1, 10, and 20 Authors by Degree Centrality**

| Field | No. of Top 20 in PGT (Threshold) | No. of Authors on or above Threshold in PGT | | | No. of Scholars who Stay in Top 10 | | | Change of Top 1 (Y/N) | | |
|---|---|---|---|---|---|---|---|---|---|---|
| | | FD | AD | HD | FD | AD | HD | FD | AD | HD |
| Biology | 62 (47) | 130 | 76 | 76 | 0 | 4 | 4 | Y | Y | Y |
| CS | 20 (35) | 112 | 32 | 39 | 0 | 3 | 3 | Y | Y | Y |
| Nanoscience | 20 (65) | 267 | 107 | 105 | 1 | 2 | 2 | Y | Y | Y |
| Neuroscience | 21 (120) | 48 | 21 | 22 | 4 | 7 | 8 | N | N | N |
| Physics | 21 (70) | 69 | 25 | 27 | 1 | 8 | 8 | Y | Y | Y |
| PNAS | 20 (78) | 126 | 36 | 43 | 0 | 3 | 3 | Y | Y | Y |

The inflation in the number of authors in the top rankings is not the only problem. Some of remaining authors in the top ranking by IBD do not have the same identity in the proxy of ground-truth data. For example, AD ranks 'Li, X. L.' (18 publication) in top 10 productive authors in computer science. Actually, 'Li, Xuelong' in the proxy of ground-truth data is ranked in top 10 with 13 publication. Five publications from three different authors ('Li, Xiaoli 1,' 'Li, Xiaoli 2,' and 'Li, Xiuli') were unduly added to the author through AD.

*Influence of Asian Names on Misidentification*

Scholars have suggested that the increase in Asian scholars, especially Chinese, Japanese and Korean scholars, contribute to name ambiguity in bibliometrics (Strotmann & Zhao, 2012; Torvik & Smalheiser, 2009). Many Asian individuals are known to share common last and/or first names, which aggravates name ambiguity. In order to find out the degree to which Asian names affect the misidentification, we calculated the proportion of 600 Asian – mostly Chinese, Indian, Japanese, Korean, Taiwan, and Vietnamese – names in misidentified author lists in each datasets. The result is summarized in the Table 10. For example, in biology coauthorship network, authors with Asian surnames constitute 17.38% of all unique authors in the network. The ratio of authors with Asian names whose identity is compromised by IBD against the total number of misidentified authors is 62.36% by FD, 69.18% by AD, and 62.84% by HD in the case of biology. Overall, it is confirmed that Asian names are highly likely to affect the name ambiguity and, thus, cause misidentification of authors across each dataset.

Table 10: Influence of Asian Names in Misidentification Per Disambiguation Method

| Field | Ratio of Unique Asian Authors in PGT | Ratio of Asian Authors among Misidentified Authors | | |
|---|---|---|---|---|
| | | FD | AD | HD |
| Biology | 17.38% | 62.36% | 69.18% | 62.84% |
| Computer Science | 31.92% | 86.72% | 84.72% | 83.00% |
| Nanoscience | 41.43% | 89.84% | 89.65% | 88.84% |
| Neuroscience | 14.63% | 54.22% | 52.06% | 52.80% |
| Physics | 20.67% | 69.17% | 61.47% | 63.03% |
| PNAS | 20.08% | 67.68% | 78.54% | 70.26% |

**Conclusion and Discussion**

In this paper, we attempted to estimate the effect of initial-based disambiguation on author identification and statistical properties of large-scale coauthorship networks. For that, we obtained articles from selected journals in five academic fields and an interdisciplinary journal where large-scale coauthorship network studies have been conducted with one of three types of IBD: first initial, all initial, and hybrid methods. Author names were disambiguated based on full name string, coauthor name, affiliation and email address in order to produce a proxy of ground-truth datasets to be compared with those disambiguated by three IBD methods.

Using these datasets, we were able to compare several statistical properties. Some properties showed an overall decrease of values in initial-based-disambiguation datasets: the number of unique authors, the number of components, average shortest paths, clustering coefficient, and assortativity. This means that when we rely on IBD we are supposed to find coauthorship networks that are smaller, where people are closer to each other, less collaborative with shared coauthors, and less homogeneous in terms of collaboration partners than they actually are. Other measures tended to increase: density, average productivity, average degree, and the proportion of the largest component. When using these techniques, scholars will appear to be more productive, collaborative and imbedded in larger and more cohesive communities than they actually are. These distortive effects of IBD were also observed in previous studies (Fegley & Torvik, 2013; Kim, Kim, & Diesner, 2014). Among other network properties, the ratio of largest component size and assortativity showed the greatest vulnerability to change by initial based disambiguation. In addition, the top authors in terms of productivity and degree centrality were changed to a non-negligible extent. Asian names were found to account for a majority of misidentified names.

We conclude that initial-based disambiguation can misidentify author identities mainly through merging, and, therefore, can distort macroscopic views of authorship pattern and collaboration structure of a field or scientific community. However, our approach has some limitations. Our datasets depended on sampled articles published in several top journals in selective fields. The selection of journals and fields might affect the details of our research findings on name ambiguity and its effects on coauthorship network measurements. In addition, as our datasets were obtained for the most recent year (2012), the distortive effect of name ambiguity might be different in datasets from another period. For example, as Asian names accounted for many of misidentification of author identities in our study, it would be possible that, in specific academic fields or time periods in which Asian names are more rare, the errors introduced by IBD can be reduced to a negligible extent. Moreover, although the distortive effect of name ambiguity is expected to deepen as the dataset increases (Fegley & Torvik, 2013), the relationship between data size and errors is still not well understood. For example, the errors may show a linear, sublinear, or sigmoid growth with the data increase. All these indicate that our findings cannot be indiscriminately generalized to other fields or studies.

Instead, the findings of our study are calling for a special attention to data pre-processing when conducting large-scale coauthorship network research targeting a field, especially where many Asian scholars participate in research. This study does not attempt to refute previous large-scale coauthorship network studies where IBD has been used. This paper may, however, serve as a nudge that, without appropriate knowledge of the effect of name ambiguity on coauthorship network properties, research findings should be accepted with care. This might be one of reasons why leading scholars such as Mark E. J. Newman and Albert-László Barabási, who have relied on IBD, recently began to perform algorithmic name disambiguation before conducting network analysis of large-scale bibliometric data (Deville et al., 2014; Martin, Ball, Karrer, & Newman, 2013)

Another important implication of this study is that naming practice and its impact on name disambiguation should be studied with a special care by bibliometric scholars. The idea that an author can be represented by a surname and initialized given names may not work for names from other regions than North America or some European countries. The proportion of scholars from different naming cultures that enter the international scholarly community is increasing. For example, the relative frequencies of Asian names in biomedical fields have been "accelerating since the mid-1980s" (Torvik & Smalheiser, 2009). From 2005 to 2011, China' share of scientific publications in science has ranked second or third following U.S. and E.U. (National Science Board, 2014). As of 2007, 30% up to 40% of assistant professors in top 100 U.S. departments in mathematics, engineering and computer science are Asians and their ratio kept increasing in various fields including social science (Nelson & Brammer, 2010). Scholars with Hispanic name origins also increase and add to the name ambiguity in bibliometrics databases. Based on these numbers we argue that names from non-traditional Western cultures should not be neglected or excluded from analysis as errors in bibliometrics.

In addition, the large portion of Asian name in misidentification in our study suggests that identification errors introduced by IBD are not randomly distributed. Studies of robustness of network measurement have mostly assumed the error of node addition/deletion or edge addition/deletion is random (Borgatti et al., 2006; Frantz et al., 2009). Our findings call for more differentiated research on the effect of name disambiguation on coauthorship network measurements. It should be also noted that authors may change names due to marriage or religion and that bibliometric data can be indexed with errors, for example, during optical character recognition. These problems have not been adequately addressed both by initial-based or algorithmic disambiguation as well as our disambiguation method, waiting for scholarly investigation of their impact on bibliometrics research.

As such, the resolution of name ambiguity in bibliometrics is not any more a matter of choice, but a necessary step in order to obtain a correct understanding of scientific collaboration, especially for a large-scale coauthorship network studies.

## Acknowledgements

This work is supported by KISTI (Korea Institute of Science and Technology Information), grant P14033, and the FORD Foundation, grant 0145-0558. We would like to thank Vetle I. Torvik and Brent D. Fegley for helpful comments on the manuscript. We also thank anonymous reviewers for their ideas that help us to improve this paper.

**Footnotes:**

1. Although there is no agreement on the definition of large-scale coauthorship networks, we assume here that a coauthorship network involving more than ten thousand authors is large-scale.

2. According to Web of Science, Newman's two works (2001, 2004) have been cited 1,641 times and Barabási et al. (2002) alone 710 times as of July 17, 2014.

3. Some scholars have cited Moody (2004) to support IBD. Moody's study used a dataset from *Sociological Abstracts* recording the majority of author names with full given names. Also, name disambiguation was discussed in the context of full given names being provided (p.219). Furthermore, the study relied on coauthor names to disambiguate target name pairs, which is thought to contribute to a high accuracy of author identification than relying only on name strings (Onodera et al., 2011; Torvik & Smalheiser, 2009). Considering these characteristics, it may be inappropriate to cite Moody (2004) for justification of IBD.

4. Milojević (2013) may be an exception to this argument. In this study, the performance of initial-based disambiguation – FD, AD, and HD - was tested through simulation based on the frequency of last names and first/middle name initials from various authorship datasets. The outcome was encouraging: the ratio of estimated "true" authors that were merged or split into other identities by FD, AD, and HD was 1.5% up to 5.5%, when only best results were considered. The study concluded that, if used in a single academic field, initial-based disambiguation can be regarded as being "quite accurate" with a few percentage of error and thus does not "have an adverse effect on many or most statistical bibliometrics studies" (p.773). However, the study relied on a *hypothesized (synthetic)* ground-truth data generated from simulation, whose accuracy was not verified, although the author stated that all the constraints gained from datasets were correctly reflected during the simulation process.

5. The list of nicknames and their corresponding full first names was obtained at http://www.usgenweb.org/research/nicknames.shtml

6. The list of Asian names was constructed as follows. First, we got the list of all names appearing at least twice across all datasets in our study. From this list, we extracted 698 Asian surnames through the name-ethnicity classification interface (Ambekar, Ward, Mohammed, Male, & Skiena, 2009). We manually corrected some errors, which resulted in a total of 650 names. The majority of them are Chinese, Indian, Japanese, Korean, Taiwan, and Vietnamese names.

7. Most of author names in our data are assigned affiliation information as Web of Science began to link the author and affiliation information staring from 2009. Some names are, however, not linked with affiliation information and others are linked with multiple affiliations. Name instances that do not have affiliation information constitute 1.33% (PNAS) to 4.40% (neuroscience) of all name instances in each of six datasets. We assign 'NA' to affiliation data field of those names. If one or both of name pair for similarity calculation have 'NA' for affiliation information, the pair is given the affiliation similarity value of zero. Author name instances that have more than one affiliation constitute 12% (computer science) to 24% (PNAS) of all name instances in each of six datasets. We manually checked name instances with more than three affiliations (up to 1.5% of all name instances in each dataset) to see if multiple affiliations are wrongly assigned to a single author while they should be assigned to others in the byline of papers. We corrected such errors, if any.

8. As we heuristically apply findings from previous algorithmic disambiguation studies, we do not assign weights to each similarity value. Each of attribute values in our disambiguation (e.g., shared email address or one coauthor in full name format) is often regarded as a golden standard to decide the true match of name pairs in generating ground-truth data (e.g., Torvik & Smalheiser, 2009). Thus, it should be noted that the good performance of our approach, as shown in the 'Evaluation of Disambiguation' section, seems to be mainly due to the data quality (e.g., more than 90 % of all name instances have full given names), not to our simplified algorithm.

9. *M-rate* is measured as the ratio of author clusters in the proxy of ground-truth data that contain name instances belonging to other identities or fail to contain name instances that should belong to the cluster *over* the total of author clusters. If expressed in terms of 'cluster F1' metric, the *M*-rate corresponds to (1 - Cluster Recall).